\documentclass[12pt]{iopart}

\pdfoutput=1
\usepackage{iopams}
\usepackage{graphicx}
\begin{document}

\title[]{A flat band at the chemical potential of a
Fe$_{1.03}$Te$_{0.94}$S$_{0.06}$ superconductor observed by
angle-resolved photoemission spectroscopy}

\author{P.~Starowicz$^1$, H.~Schwab$^{2,3}$, J.~Goraus$^4$,
P.~Zajdel$^4$, F.~Forster$^{2,3}$, J.~R.~Rak$^1$,
M.~A.~Green$^{5,6,7}$, I.~Vobornik$^8$ and F.~Reinert$^{2,3}$}

\address{$^1$ M. Smoluchowski
Institute of Physics, Jagiellonian University, Reymonta 4, 30-059
Krak\'{o}w, Poland}
\address{$^2$ Universit\"{a}t W\"{u}rzburg, Experimentelle Physik
VII, Am Hubland, D-97074 W\"{u}rzburg, Germany}
\address{$^3$ Karlsruher Institut f\"{u}r Technologie KIT,
Gemeinschaftslabor f\"{u}r Nanoanalytik, D-76021 Karlsruhe,
Germany}
\address{$^4$ Institute of Physics, University of Silesia,
Uniwersytecka 4, 40-007 Katowice, Poland}
\address{$^5$ NIST Center for Neutron Research, National Institute
of Standards and Technology, Gaithersburg, Maryland 20899, USA}
\address{$^6$ Materials Science and Engineering, University of
Maryland, College Park MD-20742-6033, USA}
\address{$^7$ School of Physical
Sciences, University of Kent, Canterbury, Kent CT2 7NH, UK}
\address{$^8$ CNR-IOM, TASC Laboratory, SS 14, km 163.5, I-34149
Trieste, Italy} \ead{pawel.starowicz@uj.edu.pl}

\begin{abstract}
The electronic structure of superconducting
Fe$_{1.03}$Te$_{0.94}$S$_{0.06}$ has been studied by angle
resolved photoemission spectroscopy (ARPES). Experimental band
topography is compared to the calculations using the methods of
Korringa-Kohn-Rostoker (KKR) with coherent potential approximation
(CPA) and linearized augmented plane wave with local orbitals
(LAPW+LO). The region of the $\Gamma$ point exhibits two hole
pockets and a quasiparticle peak close to the chemical potential
($\mu$) with undetectable dispersion. This flat band with mainly
$d_{z^{2}}$ orbital character is formed most likely by the top of
the outer hole pocket or is an evidence of the third hole band. It
may cover up to 3 $\%$ of the Brillouin zone volume and should
give rise to a Van Hove singularity. Studies performed for various
photon energies indicate that at least one of the hole pockets has
a two-dimensional character. The apparently nondispersing peak at
$\mu$ is clearly visible for 40 eV and higher photon energies, due
to an effect of photoionisation cross section rather than band
dimensionality. Orbital characters calculated by LAPW+LO for
stoichiometric FeTe do not reveal the flat $d_{z^{2}}$ band but
are in agreement with the experiment for the other dispersions
around $\Gamma$ in Fe$_{1.03}$Te$_{0.94}$S$_{0.06}$.
\end{abstract}

\pacs{74.25.Jb, 74.70.Xa, 79.60.Bm}

\maketitle

\section{Introduction}

\begin{figure}[h]
\begin{center}
\includegraphics[width=3.3in]{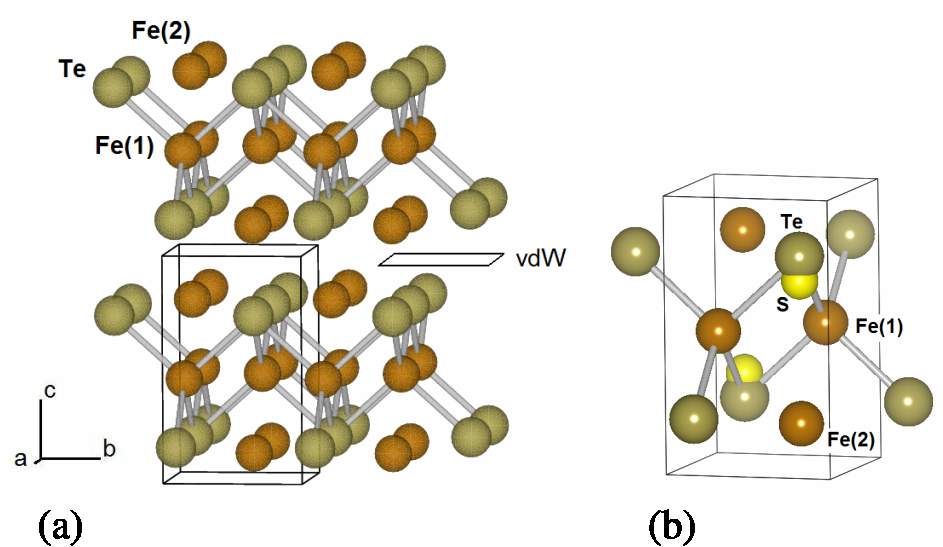}
\caption{\label{struct} (Color online) (a) Crystallographic
structure of Fe$_{1+x}$Te with marked atomic positions,
tetrahedral coordination of Fe(1) atoms, van der Waals gap (vdW),
elementary unit cell and orientation of crystallographic axes. The
atomic positions of Fe(1) and Te are completely filled, while
Fe(2) positions are only partially occupied. (b) Elementary unit
cell of sulphur doped Fe$_{1+x}$Te$_{1-y}$S$_y$ with coordination
of Fe(1) atoms. S substitutes Te but is displaced along c
crystallographic axis with respect to Te
position~\cite{Zajdel2010}. Both Te and S positions are shown.}
\end{center}
\end{figure}

The search for new superconducting materials and the opportunity
to discover further evidence of non-BCS mechanisms of electron
pairing attracted attention of researchers to iron
pnictides~\cite{Kamihara2008,Paglione2010} and
chalcogenides~\cite{Hsu2008,Fang2008,Mizuguchi2009,Mizuguchi2010,Guo2010,Krzton2011,Mizuguchi2011,Fang2011}.
Among these materials the systems from the "11" group, namely
Fe$_{1+x}$Se~\cite{Hsu2008},
Fe$_{1+x}$Te$_{1-y}$Se$_y$~\cite{Sales2009}, and
Fe$_{1+x}$Te$_{1-y}$S$_y$~\cite{Mizuguchi2009}, have the simplest
crystallographic structure with iron atoms arranged in
characteristic planes (figure \ref{struct}). These Fe(1) atoms,
tetrahedrally coordinated by chalcogen atoms, form layers
separated by van der Waals gaps. In consequence the "11" systems
can be regarded as quasi two-dimensional. Nevertheless, this
structure features an intrinsic disorder due to both excess iron
in partially occupied Fe(2) positions~\cite{Bao2009} and
substituted atoms, which are displaced with respect to the Te
crystallographic positions. It is known that doped S atoms have a
\textit{z} coordinate considerably different from that of
Te~\cite{Zajdel2010}, as displayed in figure \ref{struct} (b).

The superconducting critical temperature for "11" chalcogenides is
relatively low under ambient pressure and reaches barely 14
K~\cite{Sales2009} for Fe$_{1+x}$Te$_{1-y}$Se$_y$, 13 K for
FeSe~\cite{Hsu2008} and 10 K for
Fe$_{1+x}$Te$_{1-y}$S$_y$~\cite{Mizuguchi2009} while Fe$_{1.1}$Te
remains a non-superconducting
antiferromagnet~\cite{Mizuguchi2010}. Moreover, the
superconducting fraction of untreated FeTe$_{0.8}$S$_{0.2}$ as
determined from magnetic susceptibility is close to 20
\%~\cite{Mizuguchi2010}. The direct connection between the iron
overstoichiometry, magnetism and superconductivity can be exposed
by topotactic deintercalation using
iodine~\cite{iodineFeTe,iodineFeTeSe} or other oxidation processes
like annealing in oxygen~\cite{Mizuguchi2010a,Mizuguchi2011b}.
Samples with the lowest content of excess iron have the highest SC
fraction reaching 100\%. The promising fact is that under high
pressure the transition to superconductivity reaches T=37 K for
FeSe~\cite{Margadonna2009,Medvedev2009}.

While the mechanism of electron pairing in the Fe-based
superconductors is still under debate, the electronic band
structure can impose certain conditions on possible
scenarios~\cite{Paglione2010}. Therefore, the Fermi surface (FS)
and the electronic band structure of the discussed systems have
been extensively studied by means of angle resolved photoemission
spectroscopy (ARPES), quantum oscillations and density functional
theory (DFT) calculations~\cite{Mizuguchi2010,Paglione2010,
Richard2011}. In particular, the previous ARPES studies on "11"
chalcogenides covered both non-superconducting
Fe$_{1+x}$Te~\cite{Xia2009,Zhang2010,Liu2013} and superconducting
Fe$_{1+x}$Te$_{1+y}$Se$_y$~\cite{Nakayama2010,Chen2010,Tamai2010,Miao2012,Lubashevsky2012}
but corresponding results for FeSe or Fe$_{1+x}$Te$_{1+y}$S$_y$
are absent in the literature so far. While the published data for
Fe$_{1+x}$Te$_{1+y}$Se$_y$ are relatively consistent, studies of
Fe$_{1+x}$Te present two aspects: on the one hand clearly visible
band topography~\cite{Xia2009}, on the other hand intrinsically
broad spectra in a paramagnetic state with emergence of
quasiparticle peaks in the spin density wave (SDW)
state~\cite{Zhang2010}. The latter scenario is confirmed by a more
recent study of Fe$_{1.02}$Te and becomes understood in terms of
polaron formation~\cite{Liu2013}. The Fermi surface of
superconducting Fe$_{1+x}$Te$_{1+y}$Se$_y$ chalcogenides consists
of hole pockets located around the $\Gamma$(Z) point and electron
pockets in the region of the M(A) point, which is typical of both
iron pnictides and chalcogenides. However, the newer
A$_x$Fe$_{2-y}$Se$_2$ systems (A=K, Cs, Rb, Tl, etc.) are
exceptional in that respect as they exhibit electron pockets at
$\Gamma$(Z)point~\cite{Zhang2011,Qian2011,Mou2011,Zhao2011}.

The current paper presents the band structure and dominant orbital
characters obtained by ARPES for Fe$_{1.03}$Te$_{0.94}$S$_{0.06}$
superconductor. The data are compared to theoretical calculations.
A flat band close to the chemical potential ($\mu$) is found in
the region of the $\Gamma$ point. The resulting high density of
states at $\mu$ should be an important factor for the emergence of
superconductivity in the sulphur doped "11" compounds.

\section{Experimental}

Single crystals with targeted stoichiometry
Fe$_{1.1}$Te$_{0.875}$S$_{0.125}$ were grown in NIST by similar
techniques as reported earlier~\cite{Zajdel2010}. Stoichiometric
quantities of the elements were sealed in evacuated quartz tubes
and heated at 775 $^{\circ}$C for 48 h with intermediate step at
450 $^{\circ}$C. After regrinding the product was reheated at 825
$^{\circ}$C for 12 h and slowly cooled to room temperature. X-ray
diffraction performed at 290 K indicated single crystals with a
composition of Fe$_{1.11}$Te$_{0.91}$S$_{0.12}$ as obtained from
the Rietveld refinement to X-ray data. The determined crystal
structure as shown in figure \ref{struct} is consistent with the
previous studies~\cite{Zajdel2010} and remains tetragonal to the
lowest temperature T=35 K reached in the experiment. The
composition of the single crystals was also determined using a
JEOL JXA 8900 microprobe in wavelength dispersive mode (WDS) from
10 flat points spread over the surface. The average composition
was found to be Fe$_{1.03(1)}$Te$_{0.94(4)}$S$_{0.06(2)}$ and will
further be used in the text as more reliable than the estimate
from the diffraction data. The single crystals exhibited onset of
the superconducting transition at T=9 K in magnetic susceptibility
and electrical resistivity. However, according to the magnetic
susceptibility studies the Meissner phase at T=2K covered 23 \% of
the volume.

The ARPES experiments were carried out at the APE
beamline~\cite{Panaccione2009} of the Elettra synchrotron using
Scienta SES2002 electron spectrometer. The crystals were cleaved
at a pressure of $2\cdot10^{-11}$ mbar and studied with linearly
or circularly polarized radiation. The energy and wave vector (k)
resolution were 20 meV and 0.01 $\mathring{A}^{-1}$ respectively.
Low energy electron diffraction was used to check the surface
quality. Fermi edge determination was performed regularly on
evaporated gold.

Band structure calculations were carried out with the AkaiKKR
software~\cite{Akai} based on the Korringa-Kohn-Rostoker (KKR)
Green's function method with coherent potential approximation
(CPA). This method is able to model the effect of disorder in
alloys~\cite{Faulkner1982,Singh1994} and should treat properly
random occupancies of Fe(2) and Te/S atomic positions. CPA is
considered as the most relevant approach for disordered "11"
systems~\cite{Singh2010}. A von Barth and Hedin type
exchange-correlation potential~\cite{BarthHedin} was applied. The
width of the energy contour for the integration of the Green's
function was 1.9 Ry and the added imaginary component of energy
was 0.002 Ry. The Bloch spectral function was calculated for 255
k-points in the irreducible Brillouin zone (IBZ).

Other calculations were performed for stoichiometric FeTe by means
of the linearized augmented plane wave with local orbitals
(LAPW+LO) method implemented in the Wien2k package~\cite{Wien2k}.
Local spin density approximation (LSDA)~\cite{PerdewWang1992} and
Ceperley-Alder parametrization~\cite{CeperleyAlder1980} were used.
The atomic spheres radii were 2.41 atomic units (a.u.) and 2.17
a.u. for Fe and Te respectively, and the calculations were
realized for 330 k-points in the IBZ.

\section{Results and discussion}
\subsection{Band topography along high symmetry directions, experiment and calculations}

\begin{figure}[h]
\begin{center}
\includegraphics[width=3.4in]{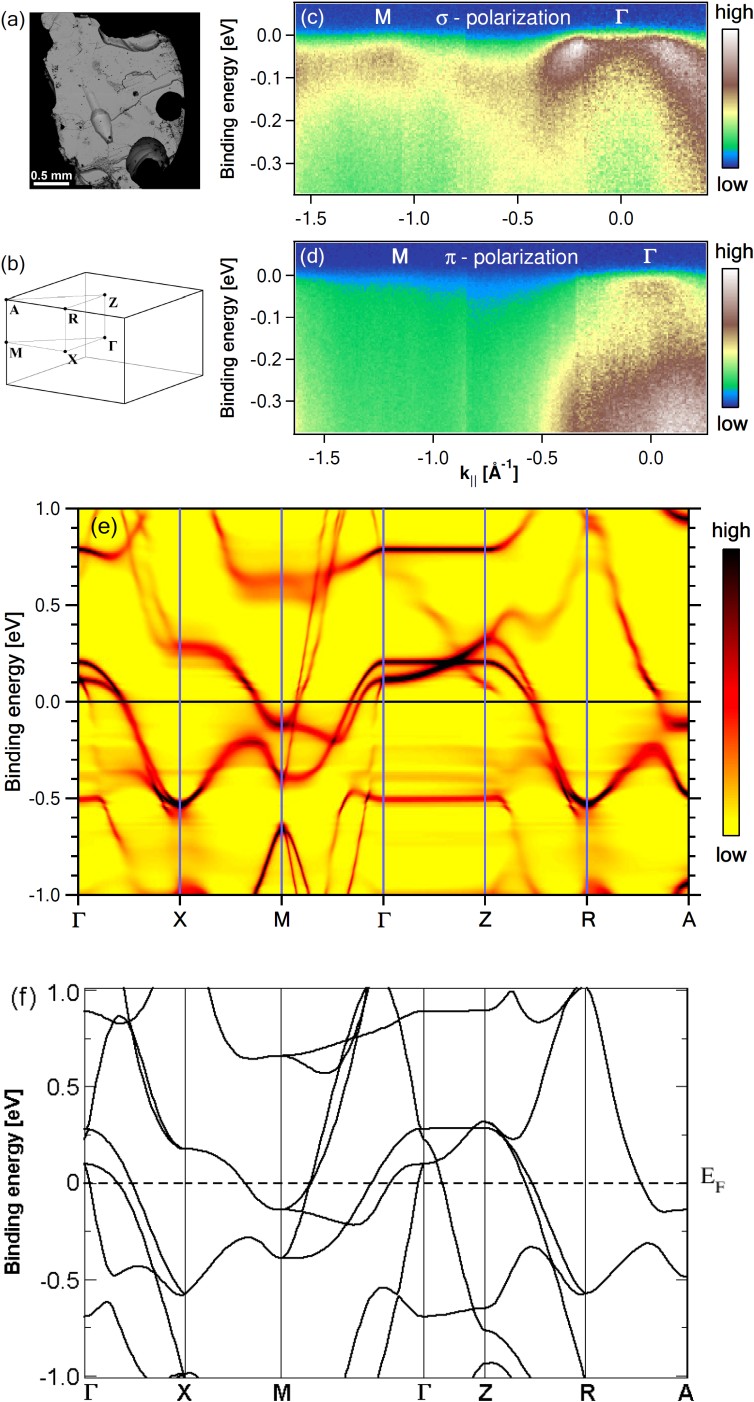}
\caption{\label{bz} (Color online) (a) Surface of
Fe$_{1.03}$Te$_{0.94}$S$_{0.06}$ single crystal exposed along
(001) plane. (b) First Brillouin zone for tetragonal
Fe$_{1+x}$Te$_{1-y}$S$_{y}$ with high symmetry points and
directions. ARPES intensity along the $\Gamma$-M (Z-A) direction
obtained at T=80K and photon energy h$\nu$=40 eV in (c)
$\sigma$-polarization and (d) $\pi$-polarization. (e) Band
structure of Fe$_{1.1}$Te$_{0.9}$S$_{0.1}$ along high symmetry
directions obtained by KKR-CPA calculations. (f) Band structure of
stoichiometric FeTe calculated with LAPW+LO method. The distances
between the high symmetry points are scaled to the real distances
in k-space (f) or remain constant between the points (e).}
\end{center}
\end{figure}

The electronic structure of superconducting
Fe$_{1.03}$Te$_{0.94}$S$_{0.06}$ crystals (figure \ref{bz}(a)) was
studied by means of ARPES along the high symmetry directions
$\Gamma$-M and $\Gamma$-X (figure \ref{bz}(b)). Radiation of
linearly polarized photons with an energy of 40 eV was used. The
spectra obtained along the $\Gamma$-M direction at 80 K (figure
\ref{bz} (c), (d)) exhibit high intensity in the region of the
$\Gamma$ point. For $\sigma$-polarization a hole pocket is found,
whereas for $\pi$-polarization the measurements reveal a hole like
band and a feature with high intensity at $\mu$. The nature of
this high spectral intensity will be discussed further.
Photoelectron spectra obtained in the region of M with
$\sigma$-polarization reveal increased intensity near $\mu$ at the
M point. The $\sigma$-polarization is more favourable for the
bands at M, similarly to the case of undoped FeTe~\cite{Xia2009}.
The spectra recorded with $\pi$-polarization do not reveal any
bands in this region. Near the X point no spectral intensity is
found at low binding energy (not shown). In particular, a replica
of the band structure at $\Gamma$ is not found at X in contrast to
the observations for undoped Fe$_{1+x}$Te~\cite{Xia2009}. This
indicates that the SDW magnetic order is not seen in the
Fe$_{1.03}$Te$_{0.94}$S$_{0.06}$ system with ARPES.

KKR-CPA calculations, which are destined for systems with
disorder, were performed for Fe$_{1.1}$Te$_{0.9}$S$_{0.1}$ (figure
\ref{bz}(e)). Despite slightly higher S content than in the
measured samples, the calculations should yield the overall effect
of doping. The theoretically obtained spectra are broadened due to
disorder, which should be reflected in the ARPES data. LAPW+LO
calculations (figure \ref{bz}(f)) were realized for stoichiometric
FeTe system as this approach cannot deal with fractional atomic
site occupancies. There is a qualitative agreement between the
band structure obtained with these two methods; in both cases
three hole pockets are present at the $\Gamma$ point, two electron
pockets are found at the M point, while there is no FS around the
X point.\underline{} The difference is observed at the M point,
where the band seen below -0.6 eV for KKR-CPA is located below
-1.2 eV for LAPW+LO results, which is out of the scale for the
figure \ref{bz}(f). Differences are also visible for the
$\Gamma$-Z direction. A dispersion along $\Gamma$-Z is a matter of
interest, as it may indicate whether the system is
two-dimensional. In fact, weak dispersions or even lack of
dispersion for certain bands are observed, what is seen in
particular for the KKR-CPA approach. This means that this system
may be considered as quasi two-dimensional to some extent. It is
also noteworthy that the dispersions near $\Gamma$ obtained with
KKR-CPA are characterized with lower slopes and higher band masses
as compared to LAPW+LO at low binding energies. The discrepancies
between the obtained band structures may have arisen from
different exchange-correlation potential and different modeling of
atomic spheres in the approaches as well as due to the differences
between the objects of the studies; Fe$_{1.1}$Te$_{0.9}$S$_{0.1}$
and FeTe. To obtain the agreement between the experiment and the
theory the Fermi energy for the calculated band structure needs to
be shifted up by 0.11 eV and 0.10 eV for KKR-CPA and LAPW+LO
respectively.

The band structure obtained from the calculations is generally
consistent with the ARPES results both along the $\Gamma$-M and
$\Gamma$-X directions assuming that certain bands may be invisible
in the experiment due to unfavourable matrix elements. Out of the
three hole pockets predicted by calculations at least two
hole-like bands at $\Gamma$ are found in the experiment.
Theoretical results are also consistent with the spectra near M
taken along the $\Gamma$-M direction (figure \ref{bz}(c)), where a
band moves towards $\mu$ when k approaches M, which is visible for
$\sigma$-polarization. The calculated electron pocket at M is not
resolved in the experiment. Theoretical dispersions along
$\Gamma$-X confirm the absence of energy bands near $\mu$ at X.

\subsection{Band structure near the $\Gamma$ point}

\begin{figure}[h]
\begin{center}
\includegraphics[width=6.2in]{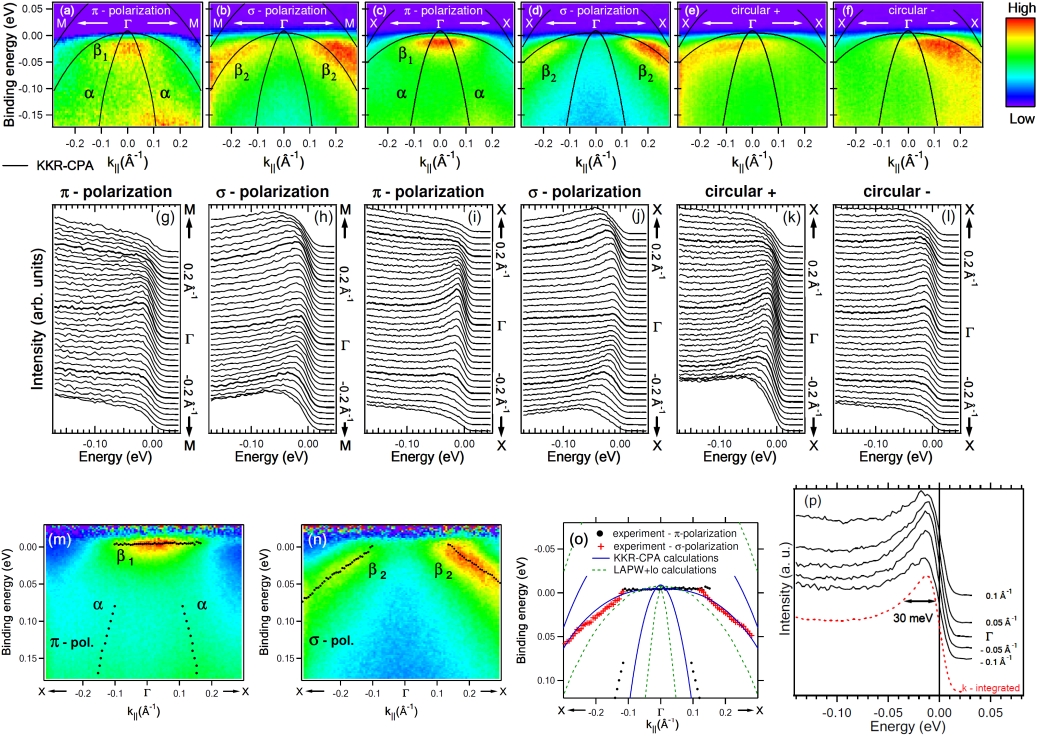}
\caption{\label{gamma} (Color online) Energy bands in the $\Gamma$
point region for Fe$_{1.03}$Te$_{0.94}$S$_{0.06}$ obtained by
ARPES along the M-$\Gamma$-M direction with (a) $\pi$ and (b)
$\sigma$ polarizations and along X-$\Gamma$-X, with (c) $\pi$, (d)
$\sigma$, (e) circular plus and (f) circular minus polarizations.
The experimental dispersions are named as $\alpha$, $\beta_{1}$
and $\beta_{2}$. Theoretical dispersions obtained by KKR-CPA
calculations (solid lines) are superimposed on the graphs. The
spectra are shown as energy distribution curves (EDCs) in (g) -
(l). The spectra from (c) and (d) divided by the Fermi function
are presented in (m) and (n) respectively. Experimental band
dispersions marked by black points result from energy or momentum
distribution curve fitting. The extracted dispersions are compared
to KKR-CPA (for Fe$_{1.1}$Te$_{0.9}$S$_{0.1}$) and LAPW+LO (for
FeTe) calculations (o). Panel (p) shows extracted EDCs from (c)
[or (i)] and the curve resulting from wave vector (k) integration
of all EDCs between $-0.1 \mathring{A}^{-1}$ and $0.1
\mathring{A}^{-1}$ from (c) [or (i)] - red line (dashed). All
measurements were performed with incident photon energy of 40 eV
at the temperature of 35 K.}
\end{center}
\end{figure}

Let us analyze the region of the $\Gamma$ point for
Fe$_{1.03}$Te$_{0.94}$S$_{0.06}$, where the band structure appears
to be different from that observed before for undoped
non-superconducting Fe$_{1+x}$Te~\cite{Xia2009,Zhang2010,Liu2013}.
ARPES studies performed at T=35 K include scans along M-$\Gamma$-M
with $\pi$ and $\sigma$ polarizations as well as along
X-$\Gamma$-X with $\pi$, $\sigma$, circular plus and circular
minus polarizations (figure \ref{gamma} (a)-(l)). Solid lines
representing dispersions from KKR-CPA calculations (figure
\ref{bz}(e)) are drawn on the experimental data in figure
\ref{gamma} (a) - (f). They should be treated as guides to the eye
as they are the results of fitting to the intensity map of KKR-CPA
calculations. In order to trace the dispersions in the vicinity of
$\mu$ the spectra were divided by the Fermi-Dirac distribution and
are shown in figure \ref{gamma}(m) and (n) with binding energies
determined from fitting energy distribution curves (EDCs) or
momentum distribution curves (MDCs) with the Lorentzian function.
The experimental and theoretical dispersions are compared in
figure \ref{gamma} (o).

A comparison of the band dispersions measured along the $\Gamma$-M
(figure \ref{gamma} (a),(b),(g),(h)) and $\Gamma$-X (figure
\ref{gamma} (c)-(f), (i)-(l)) yields that they are quite similar
at the $\Gamma$ point. For $\pi$-polarization a barely visible
inner hole like band ($\alpha$) (figure \ref{gamma} (m)) can be
traced in both directions. The same polarization also yields a
very flat quasiparticle band with strong intensity near the
$\Gamma$ point ($\beta_{1}$). In fact, due to its high effective
mass the dispersion was not measurable and the band exhibits
practically constant binding energy determined to be 3 - 5 meV
above the Fermi level. The negligibility of the dispersion was
confirmed by EDCs shown in figure \ref{gamma} (p), which have
approximately the same shape at $\Gamma$ and at $\pm$0.05
$\mathring{A}^{-1}$. EDCs from $\pm$0.1 $\mathring{A}^{-1}$ at the
edges of $\beta_{1}$ seem to be more complex. Their coherent part
has the same binding energy but exhibits lower intensity. A
contribution from another structure at higher binding energy is
also observed. This structure may be evidence of a broadening of
the quasiparticle band, incoherent spectral intensity or another
hole band. Integrating the EDCs in the range $\pm$ 0.1
$\mathring{A}^{-1}$ over wave vector yields a peak with a width of
30 meV shown in figure \ref{gamma} (p). This narrow width, which
is also characteristic of single EDCs confirm the quasiparticle
nature of this spectral intensity. Raising the temperature to 70 K
did not deliver any evidence of electron like dispersion (not
shown). On the other hand, the spectra obtained with
$\sigma$-polarization (figure \ref{gamma}(b), \ref{gamma}(d),
\ref{gamma}(h), \ref{gamma}(j)) show a dispersion ($\beta_{2}$),
which looks like the outer hole pocket.

It is rather clear that $\alpha$ corresponds to the inner
hole-like band in the calculations. However, the interpretation of
$\beta_{1}$ and $\beta_{2}$ leaves certain ambiguity. The favoured
scenario assumes that these features originate from the same band.
This is supported by the circular polarization studies, which
yield a continuous dispersion of $\beta_{1}$ and $\beta_{2}$.
Moreover, such an interpretation is in agreement with the band
structure calculations (figure \ref{gamma}(o)) as $\beta_{1}$ and
$\beta_{2}$ match well the calculated middle hole band. However it
has to be remarked that the experimental dispersion exhibits a
more "kink-like" shape with mass renormalization near $\mu$ when
compared to the theoretical one. It is noteworthy that this band
changes its orbital character rather abruptly around the $\Gamma$
point, as $\beta_{1}$ and $\beta_{2}$ are sensitive to different
polarizations in the experiment. One may still consider the other
interpretation. The hypothesis that $\alpha$, $\beta_{1}$ and
$\beta_{2}$ originate from three hole pockets, can also be
compatible with our data. It may be supported by a possible
similarity between S doped and Se doped Fe$_{1+x}$Te. The band
structure at $\Gamma$ found in FeTe$_{1-y}$Se$_{y}$
before~\cite{Chen2010,Tamai2010,Miao2012} consists of three hole
like bands. In the case of
FeTe$_{0.55}$Se$_{0.45}$~\cite{Miao2012} one of the bands forms
also a flat dispersion near $\mu$ with a narrow quasiparticle
peak. An extension of this band is visible as a hole pocket.
However, in our case, the hypothesis that $\beta_{1}$ originates
from the third hole pocket, is not indicated directly by the data.
Importantly and independently of the interpretation $\beta_{1}$
remains flat and lies close to the chemical potential on a circle
with a radius of approximately 0.15 $\mathring{A}^{-1}$. Such a
situation should result in a spike in the density of states close
to $\mu$ called Van Hove singularity (VHs). It is known as an
important factor for induction or enhancement of
superconductivity. It has been already suggested that VHs may play
an important or even more universal role in the formation of
superconductivity~\cite{Borisenko2010} for a number of compounds.

Let us compare the spectra obtained for superconducting
Fe$_{1.03}$Te$_{0.94}$S$_{0.06}$ near the $\Gamma$ point with the
literature results for undoped
Fe$_{1+x}$Te~\cite{Xia2009,Zhang2010,Liu2013}. The bands found
with $\sigma$-polarization by Xia et al. \cite{Xia2009} are in
relative agreement with our spectra. However, for
$\pi$-polarization, the spectrum of Fe$_{1+x}$Te consists of a
hole pocket with no trace of the flat band at $\mu$. On another
hand the ARPES studies of undoped Fe$_{1.06}$Te~\cite{Zhang2010}
and Fe$_{1.02}$Te~\cite{Liu2013} are characterized by broadened
spectra with less clear band topography, which may be
similar~\cite{Zhang2010} or rather different~\cite{Liu2013} from
Fe$_{1.03}$Te$_{0.94}$S$_{0.06}$ results.

It is known that bands in Fe$_{1+x}$Te$_{1-y}$Se$_{y}$ appear to
be strongly renormalized~\cite{Nakayama2010,Chen2010,Tamai2010}
when compared to ab-initio calculations. In the case of
Fe$_{1.03}$Te$_{0.94}$S$_{0.06}$ the inner hole pockets from
KKR-CPA calculations fit the experimental spectra quite reasonably
(figure \ref{gamma}(o)) and do not indicate strong mass
renormalization. However, if the hypothesis of three hole pockets
in the experiment was assumed, the agreement between the data and
the calculations would be poorer. It is noteworthy that KKR-CPA
calculations made for disordered Fe$_{1.1}$Te$_{0.9}$S$_{0.1}$ and
LAPW+LO calculations performed for stoichiometric FeTe reveal
different effective masses at $\mu$ (figure \ref{gamma}(o)). This
result shows that the estimation of band renormalization can be
uncertain, as it depends on the used approach in band structure
calculations. The KKR-CPA approach yields higher effective mass in
the theoretical dispersions, what implies lower mass
renormalization.

\subsection{Photon energy dependent studies}

\begin{figure}[h]
\begin{center}
\includegraphics[width=3.2in]{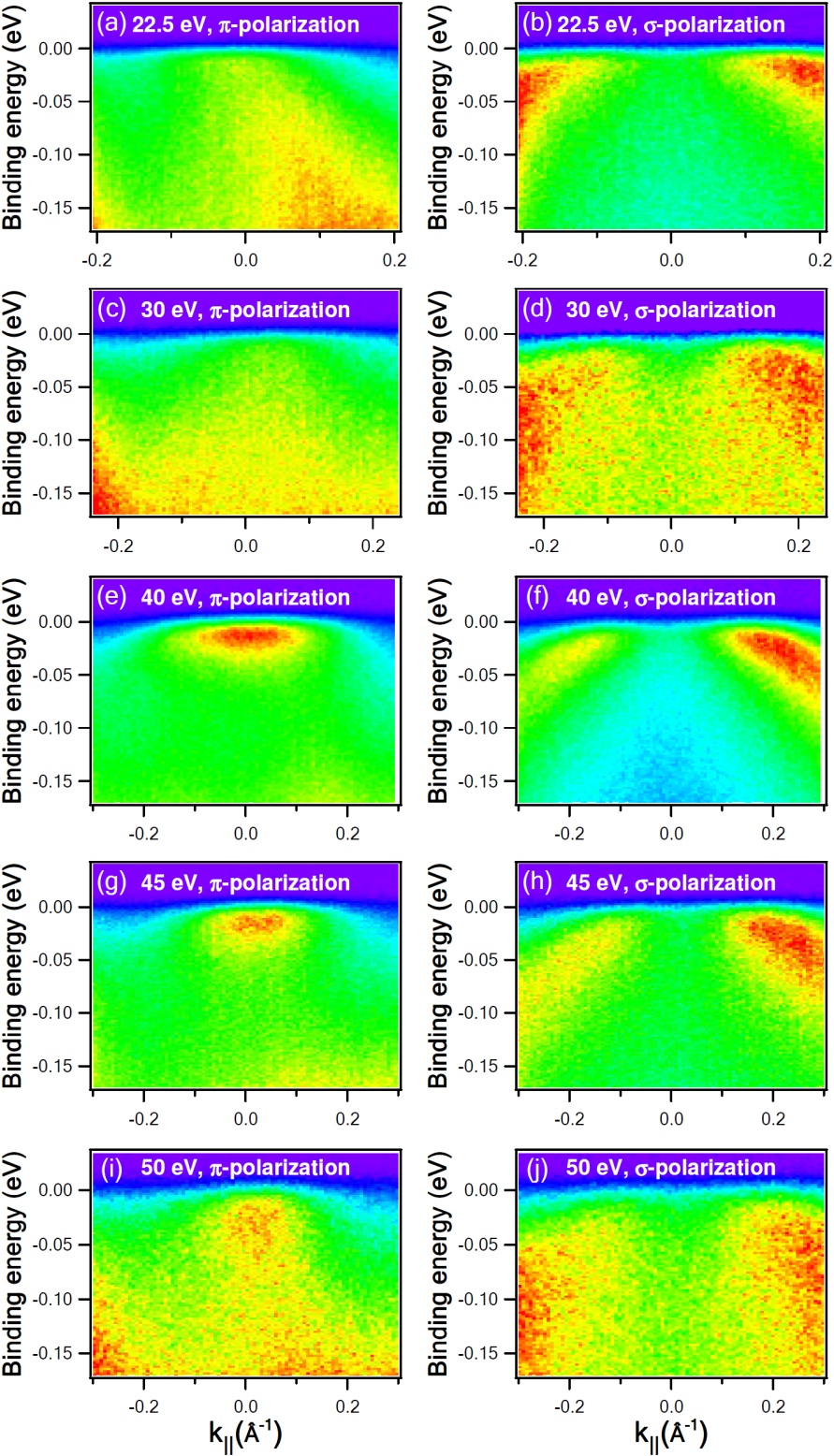}
\caption{\label{photon} (Color online) Incident photon energy
dependence of ARPES spectra recorded for
Fe$_{1.03}$Te$_{0.94}$S$_{0.6}$ at T=35 K along X-$\Gamma$-X in
the center of Brillouin zone with the following photon energies
h$\nu$ and polarizations: (a) 22.5 eV, $\pi$, (b) 22.5 eV,
$\sigma$, (c) 30 eV, $\pi$, (d) 30 eV, $\sigma$, (e) 40 eV, $\pi$,
(f) 40 eV, $\sigma$, (g) 45 eV, $\pi$, (h) 45 eV, $\sigma$, (i) 50
eV, $\pi$, (j) 50 eV, $\sigma$.}
\end{center}
\end{figure}

The next important point is band dimensionality, which can be
explored by a photon energy dependent study. Therefore, the region
of $\Gamma$ was investigated with energies between 22.5 eV and 50
eV (figure \ref{photon}). The outer part of the hole pocket
($\beta_{2}$) can always be detected with $\sigma$-polarization.
The flat dispersion near $\Gamma$ ($\beta_{1}$) can be seen for
photon energies of 40 eV, 45 eV and 50 eV. On the other hand, its
intensity is suppressed for 22.5 eV and 30 eV. There are two
optional explanations for this fact: a dispersion along the wave
vector component perpendicular to the surface ($k_z$) or a
photoionization cross section effect. To estimate the change of
$k_z$ for the considered photon energy range one may use the free
electron final state (FEFS) model~\cite{Huefner} with a typical
value of $V_0$=15 eV for the inner potential estimated in a case
of iron pnictides~\cite{Vilmercati2009,Xu2011}. If the photon
energy is increased from 22.5 eV to 50 eV the corresponding shift
in $k_z$ would be 1.06 $\mathring{A}^{-1}$, which is approximately
equal to the lattice constant in the reciprocal space c*=1.02
$\mathring{A}^{-1}$. An assumption of different $V_0$ values
between 10 eV and 25 eV does not change the corresponding shift in
$k_z$ considerably. Therefore, if the FEFS model is applicable,
the spectra for 22.5 eV and 50 eV should refer to equivalent
regions in the reciprocal lattice. In such a case different matrix
elements could be the only explanation for the vanishing spectral
intensity ($\beta_{1}$) for lower photon energies. If the flat
band is present for all $k_z$ values, it can be estimated that it
covers about 3 $\%$ of the Brillouin zone volume. Finally,
eventual dispersion of $\beta_{2}$ as a function of $k_z$ was not
found, so this band can be considered as two-dimensional.

\subsection{Analysis of the orbital characters}

\begin{figure}[h]
\begin{center}
\includegraphics[width=4.2in]{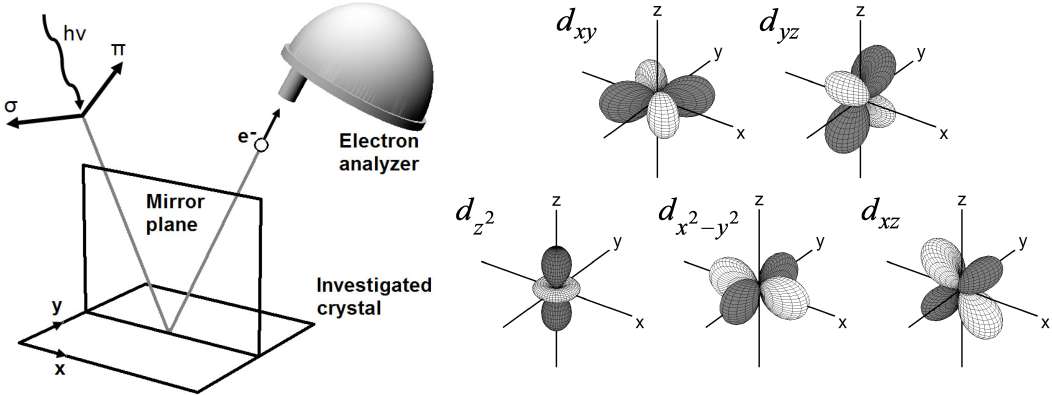}
\caption{\label{experiment} Schematic presentation of the ARPES
experiment with $\pi$-polarized photons (electric field vector in
the mirror plane) and $\sigma$-polarized photons (electric field
vector perpendicular to the mirror plane). For the sketched
configuration $\sigma$ polarized photons detect d$_{xy}$ and
d$_{yz}$ orbitals whereas $\pi$-polarized radiation probes
d$_{z^{2}}$, d$_{x^{2}-y^{2}}$ and d$_{xz}$ orbitals.}
\end{center}
\end{figure}

A photoelectron spectroscopy experiment realized in $\sigma$ or
$\pi$ geometry is able to determine the orbital wave function
parity with respect to the mirror plane, which is defined by the
positions of radiation source, sample and detector (figure
\ref{experiment})~\cite{Damascelli2004,Thirupathaiah2010,He2010,Xia2009}.
Thus, possible orbital characters can be associated with the
observed bands shown in figure \ref{gamma}. In the first
considered geometry the mirror plane is defined by the \textit{z}
axis perpendicular to the sample surface and the \textit{x} axis
corresponding to the $\Gamma$-M direction. The analyzer slit is
oriented along this plane. The orientation of the Fe - d orbitals
dominating the vicinity of the Fermi energy is similar to the case
of the iron pnictides~\cite{Singh2009} with the \textit{x} and
\textit{y} axes pointing along corresponding $\Gamma$-M
directions. $\pi$-polarized photons excite the states that are
even with respect to the considered plane. Consequently, the
$d_{xz}$, $d_{x^{2}-y^{2}}$ and $d_{z^{2}}$ orbitals are allowed
for the band $\alpha$ along $\Gamma$-M (figure \ref{gamma} a, g).
$\beta_{1}$ will be discussed separately as a special case related
to the $\Gamma$ point, which was scanned four times with different
geometries and polarizations. $\sigma$-polarized radiation probes
states with $d_{yz}$ and $d_{xy}$ orbital character, as they are
odd with respect to the mirror plane (figure \ref{experiment}).
Hence, $\beta_{2}$ along $\Gamma$-M (figure \ref{gamma} b,h) may
be dominated by these orbital characters. A rotation of the sample
such that the mirror plane is along the $\Gamma$-X direction
changes the orbital parity related to the plane. Along this
direction the orbitals $d_{xz}$ and $d_{yz}$ equally contribute to
bands as $d_{xz}+d_{yz}$ or $d_{xz}-d_{yz}$. In this geometry
measurements with $\pi$-polarization (figure \ref{gamma} c,i)
probing the bands with even symmetry indicate that $\alpha$ can be
dominated by $d_{xz}+d_{yz}$, $d_{z^{2}}$ and $d_{xy}$. On the
other hand the experiment with $\sigma$-polarization (figure
\ref{gamma} d,j) reveals that $\beta_{2}$ should originate from
$d_{x^{2}-y^{2}}$ and $d_{xz}-d_{yz}$ along $\Gamma$-X.

Finally, let us consider the $\beta_{1}$ spectrum. Bands scanned
along $\Gamma$-M with $\pi$-polarization (figure \ref{gamma}
(a,g)) can be composed of $d_{xz}$, $d_{x^{2}-y^{2}}$ and
$d_{z^{2}}$. However, the same $\Gamma$ point is also scanned
along $\Gamma$-X with $\sigma$-polarization (figure \ref{gamma}
(d,j)). The later measurement yields no intensity at $\Gamma$ what
indicates that $d_{x^{2}-y^{2}}$ and $d_{xz}-d_{yz}$ band
characters are not present there. Hence, only the $d_{z^{2}}$
remains as a dominant character for $\beta_{1}$. Similar reasoning
for the $\Gamma$ point may be done using the spectra obtained with
$\pi$-polarization along $\Gamma$-X (figure \ref{gamma} (c,i))
permitting $d_{xz}+d_{yz}$, $d_{z^{2}}$ and $d_{xy}$ characters
together with the other scan with $\sigma$-polarization along
$\Gamma$-M (figure \ref{gamma} (b,h)) revealing the lack of
intensity at $\Gamma$. The last one indicates that $d_{yz}$ and
$d_{xy}$ are not present at $\Gamma$, what leads to the same
conclusion that mainly $d_{z^{2}}$ character contributes to the
$\beta_{1}$ spectrum.

\begin{figure}[!]
\begin{center}
\includegraphics[width=6.1in]{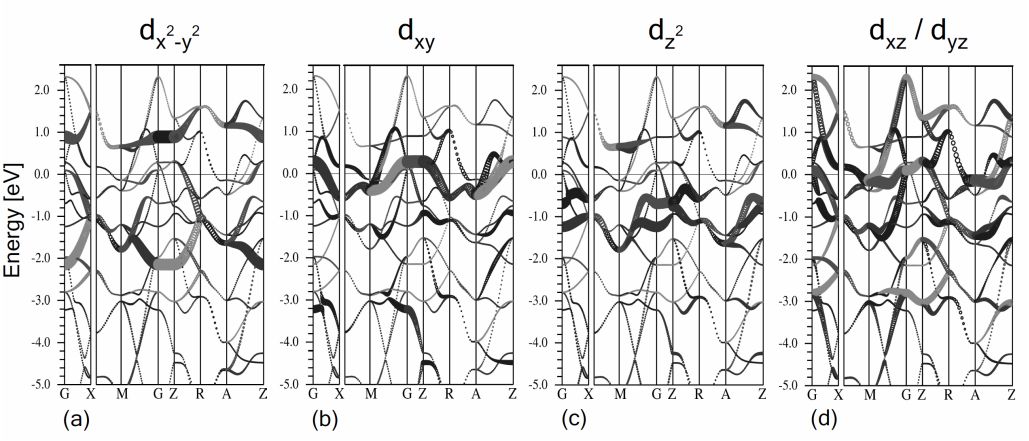}
\caption{\label{bands} Band structure of stoichiometric FeTe.
Contributions of (a) d$_{x^{2}-y^{2}}$, (b) d$_{xy}$, (c)
d$_{z^{2}}$ and (d) d$_{xz}$/d$_{yz}$ orbital characters are
represented by band widths (fat bands).}
\end{center}
\end{figure}

The contribution of s-, p- and d- valence orbital characters was
also estimated theoretically by means of LAPW+LO method
implemented in the Wien2k package~\cite{Wien2k} (figure
\ref{bands}). The calculations were realized for stoichiometric
FeTe. The results confirm that d-orbitals dominate the band
structure in the vicinity of the Fermi energy (other orbital
projections are not shown in figure \ref{bands}). The hole bands
$\alpha$ and $\beta_{2}$ appearing around the $\Gamma$ point have
their counterparts in the theoretical results. Although it is not
obvious to what extent the calculations for pure FeTe are reliable
for Fe$_{1.03}$Te$_{0.94}$S$_{0.06}$, they can narrow down the
list of possible band characters. The calculations yield that the
$\alpha$ band has mainly $d_{xz}/d_{yz}$ orbital character, while
$\beta_{2}$ is dominated by $d_{xz}/d_{yz}$ and $d_{xy}$ with some
contribution of $d_{x^{2}-y^{2}}$ along $\Gamma$-X. This is in
agreement with the experimental results obtained both along
$\Gamma$-M and $\Gamma$-X directions. In contrast, the
calculations for FeTe do not reveal the flat band at the Fermi
energy with dominant $d_{z^{2}}$ orbital character, which would
correspond to $\beta_{2}$. In this aspect they are not compatible
with the experiment for Fe$_{1.03}$Te$_{0.94}$S$_{0.06}$. One may
expect that S doping in Fe$_{1+x}$Te$_{1-y}$S$_{y}$ system may
have a particular effect on the $d_{z^{2}}$ orbital as it results
in shrinking the $c$ lattice constant.

\section{Conclusions}

The band structure of superconducting
Fe$_{1.03}$Te$_{0.94}$S$_{0.06}$ was studied along the $\Gamma$-X
and $\Gamma$-M directions by ARPES. An increased spectral
intensity at $\mu$ is observed near the $\Gamma$ and M points. In
particular, two hole bands ($\alpha$ and $\beta_{2}$) are found
around $\Gamma$ with a high intensity quasiparticle peak
($\beta_{1}$) located close to $\mu$, with no evidence of
dispersion. This latter feature has mainly $d_{z^{2}}$ orbital
character and is interpreted as the maximum of the $\beta_{2}$
hole band or an evidence of another hole pocket. Such a band
structure yields a high density of states at the chemical
potential, interpreted as a Van Hove singularity. Measurements
performed with variable photon energy show no dispersion of the
$\beta_{2}$ hole band as a function of $k_z$. Hence, it is
considered as two dimensional. The flat part of the band located
at $\mu$ has a reduced intensity for the photon energies of 30 eV
and 22.5 eV, which is attributed to a low photoionization
cross-section. The band structure obtained from KKR-CPA
calculations includes the broadening due to disorder and exhibits
three hole pockets in $\Gamma$ and two electron pockets at M.
Further LAPW-LO calculations performed for stoichiometric FeTe
lead to a band topography, which is in reasonable agreement with
the KKR-CPA results and the experiment for
Fe$_{1.03}$Te$_{0.94}$S$_{0.06}$. The orbital characters
calculated with the LAPW-LO method agree with the experimental
results for $\alpha$ and $\beta_{2}$ dispersions but are
inconsistent with the $d_{z^{2}}$ character observed for the flat
$\beta_{1}$ spectrum.

\section{Acknowledgments}

Some authors (H.S., F.F. and F.R.) acknowledge the support by the
DFG through the FOR1162. The study has been supported by Polish
National Science Centre grant 2011/01/B/ST3/00425. P.Z.
acknowledges use of the equipment at the UMD Nanoscale Imaging
Spectroscopy and Properties Laboratory. The research leading to
these results has received funding from the European Community's
Seventh Framework Programme (FP7/2007-2013) under grant agreement
number 226716.


\end{document}